\documentclass{article}
\usepackage{fleqn,ams}
\usepackage{amssymb}
\usepackage{epsfig}
\usepackage{verbatim}

\def\stackunder#1#2{\mathrel{\mathop{#2}\limits_{#1}}}
\begin{document}

\begin{center}
{\bf\large Statistical systems of particles with scalar
interaction in cosmology}\\[12pt]
Yu.G. Ignat'ev, R.FA. Miftakhov\\
Kazan State Pedagogical University,\\ Mezhlauk str., 1, Kazan
420021, Russia
\end{center}

\begin{abstract}
Cosmological solutions of the equations with scalar interaction
are being studied. It is shown, that the scalar field can
effectively change the equation of state of statistical system,
that leads to series of cosmological consequences.\end{abstract}

\section {Statistical systems of particles with scalar interaction}
Statistical systems of particles with scalar interaction were
considered for the first time by one of the Authors. In
particular, in Ref. \cite{Yukin1}, \cite{Yukin2} is shown, that
the correct insertion of inter\-par\-tic\-le scalar interaction in
the kinetic theory leads to the change of particles effective
masses:
\begin{equation}\label{0}m_*=|m + q\Phi|, \end{equation}
where $q$ - a scalar charge of
particles, $\Phi$ - potential of scalar field.

\subsection {Hamilton description formalism of the particles movement
in the scalar field}

The canonical equations of the relativistic particles motions
concerning to the pair of canonically conjugated dynamic variables
$x^i $ (coordinates) and $P_i $ (the ge\-ne\-ra\-li\-zed momentum)
take the form of (see, for example, \cite {Yukin1}):
\begin{equation} \label {1}
\frac {d x^i} {ds} = \frac {\partial H} {\partial P_i}; \qquad
\frac {d P_i} {ds} = - \frac {\partial H} {\partial x^i},
\end{equation}
where $H (x, P) $ -relativistically invariant Hamilton function.
Calculating full derivative of the dynamic variables function
$\Psi (x^i, P_k) $, in view of (\ref {1}) we shall find:
\begin{equation} \label {2}
\frac {d \Psi} {ds} = [H, \Psi],
\end{equation}
where invariant Poisson brackets are introduced:
\begin{equation} \label {3}
[H, \Psi] = \frac {\partial H} {\partial P_i} \frac {\partial
\Psi} {\partial x^i} - \frac {\partial H} {\partial x^i} \frac
{\partial \Psi} {\partial P_i}.
\end{equation}
In consequence of (\ref {3}) Hamilton function is the integral of
particle movements, - this integral of movement is called the
rest-mass of particles:
\begin{equation} \label {4}
\frac {d H}{ds}=[H,H]=0, \Rightarrow H = \frac {1} {2} m^2 = \mbox
{Const}.
\end{equation}
The relation (\ref {4}) is called the relation of normalization.
Relativistically-invariant Hamilton function particles with scalar
charge $q $, being in a scalar field with potential $\Phi $, is
\cite{Yukin1}:
\begin{equation} \label {4}
H (x, P) = \frac {1} {2} m \left [\frac {(P, P)} {m+q \Phi}-q \Phi
\right],
\end{equation}
where $ (a, b) =g _ {ik} a^ib^k $-scalar product.

The usual kinematic particle momentum, $p^i $, is determined
through Hamilton function by means of the relation:
\begin{equation} \label {5}
p^i = \frac {\partial H} {\partial P_i} \quad (= \frac {dx^i}
{ds}).
\end{equation}
Hence, taking into account (\ref {4}),  we'll receive connection
between generalized and kinematic particle momentums:
\begin{equation} \label {6}
p^i =\frac {P^i}{{\displaystyle 1 + \frac {q \Phi} {m}}}.
\end{equation}
 In consequence of (\ref {4}) following relations of momentum normalization are carried out:
\begin{equation} \label {7}
(P, P) = (m+q \Phi)^2; \Rightarrow (p, p) =m^2,
\end{equation}
whence it follows that the generalized $P_i $ and kinematic $p^i $
momentums are timelike vectors; in further we'll term the value
\begin{equation} \label {8}
m_ * = m+q \Phi
\end{equation}
as an effective mass of a particle.  Its meaning is shown in
equations of motion (\ref {1}), which concerning to Hamilton
function (\ref {4}) become:
$$
\left(1+\frac{q\phi}{m}\right)\left[\frac{d^2
x^i}{ds^2}+\Gamma^i_{jk}\frac{dx^j}{ds}\frac{dx^k}{ds}\right]=
$$
\begin{equation}\label{9}
=q\Phi,_k \left(g^{jk}-\frac{1}{m^2}\frac{dx^j}{ds}\frac{d x^k}{ds
}\right).
\end{equation}
The right side (\ref {9}) is a 4-vector of force $F^i $, forcing
on the particle from the direction of scalar field orthogonally to
its velocity vector:

\begin{equation} \label {10}
(F, p) =0.
\end{equation}
Conversion to a usual normalization of velocity vector
$$ u^i = \frac{dx^i}{ds_0}, $$
so that
$$ (u, u) =1, $$
is proceeded by re\-nor\-ma\-li\-za\-ti\-on of an interval:
\begin{equation} \label {11}
s = \frac {s_0} {m}.
\end{equation}
In terms $s_0 $ equations of motion take more usual form:
$$\left(1+\frac{q\phi}{m}\right)\left[\frac{d^2
x^i}{ds_0^2}+\Gamma^i_{jk}\frac{dx^j}{ds_0}\frac{dx^k}{ds_0}\right]=$$
$$ =q\Phi,_k \left(g^{jk}-\frac{dx^j}{ds_0}\frac{d x^k}{ds_0
}\right).$$

At availability of space and field symmetry along the vector field
$\xi^i (x) $ (Killing field)
\begin{equation} \label {11a}
\stackunder {\xi} {L} g _ {ik} =0; \qquad \stackunder {\xi} {L}
\Phi=0,
\end{equation}
where $\stackunder{\xi}{L}\Psi $ - Lie derivative of object $\Psi$
along a vector field $ \xi $, the canonical equations of motion
(\ref {1}) or (\ref {11}) admit linear integrals of motion (see,
for example, \cite {Eq}, \cite {Yu2}):
\begin{equation} \label {11b}
\varphi = (\xi, P) = \mbox {Const},
\end{equation}
that make sense of energy and moment integrals.

\subsection{Distribution function and macroscopic currents of
systems of particles with scalar interaction} Let $F (x, P) $ -
invariant distribution function of particles in a 8-dimensional
phase space. We shall define the momentums concerning distribution
$F (x, P) $ \cite {Yukin3}:
\begin{equation} \label {12}
n^i (x)= \int \limits _ {P (x)} p^idP,
\end{equation}
number density vector of particles, so that:
\begin{equation} \label {13}
n^i=n v^i,
\end{equation}
where $v^i $-timelike identity  vector of kinematic macroscopic
velocity of particles:
\begin{equation} \label {14} n = \sqrt {(n, n)};
\end{equation}
and
\begin{equation} \label {15}
T^{ik} _p (x) = \left (1 + \frac {q \phi} {m} \right) \int \limits
_ {P (x)} F (x, P) p^ip^kdP,
\end{equation}
- macroscopic momentum-energy tensor (MET) of particles in the
universal units system ($G = \hbar=c=1 $). An invariant element of
volume of 4-dimensional momentum space in expressions (\ref {12}),
(\ref {15}) in same units is:
\begin{equation} \label {16}
dP = \frac {2S+1} {(2 \pi) ^3 \sqrt {-g}} dP_1 \wedge dP_2 \wedge
dP_3 \wedge dP_4.
\end{equation}
The invariant 8-dimensional  distribution function \\$F(x,P)$ is
connected with 7-dimensional dis\-t\-ri\-bu\-ti\-on function $f(x,
P) $ using $\delta $-functions by relationship:
\begin{equation} \label {17}
F (x, P) =f (x, P) \delta (H - \frac {1} {2} m^2).
\end{equation}
Solving (\ref {16}) with the plus value of an energy, we shall
obtain an invariant volume unit of 3-dimensional momentum space:
\begin{equation} \label {18}
dP _ + = \left|1 + \frac {q \Phi} {m} \right | \frac {2S+1} {(2
\pi) ^3 \sqrt {-g}} \frac {dP_1 \wedge dP_2 \wedge dP_3} {|P^4 |},
\end{equation}
or in terms of kinematic momentum:
\begin{equation}\label{19}
dP_+=\left|1+\frac{q\Phi}{m}\right|^3\frac{(2S+1)\sqrt{-g}}{(2\pi)^3}\frac{dp^1\wedge
dp^2\wedge dp^3}{|p^4|}.
\end{equation}
The sign of absolute value in the above mentioned expressions
appears as a result of properties of $ \delta $-functions. It is
necessary to note, that the form (MET) (\ref {15}), received for
scalar charged particles in Ref.\cite {Yukin3}, is the unique
consequence of the supposition about total momentum conservation
during local particles collisions:
$$ \sum \limits _ {I} P_i = \sum \limits _ {F} P_k, $$
where summing is carried out by all initial and final states.

Let following responses flow past in system:
\begin{equation} \label {20-1}
\sum \limits _ {B=1}^{m '} \nu ' _Ba ' _B \rightleftarrows \sum
\limits _ {A=1} ^m \nu_A a_A,
\end{equation}
where $a_A $ - symbols of particles, and $ \nu_A $ - their
numbers. Then
$$P_I=\sum\limits_{B=1}^{m'}\sum\limits_{\alpha'}^{\nu'_B} P^{\alpha'}_B,$$
$$P_F=\sum\limits_{A=1}^m\sum\limits_\alpha^{\nu_A} P^\alpha_A,$$
and  distribution functions of particles  are described by
invariant kinetic equations \cite{Yukin3}:
\begin{equation} \label {20}
[H_a, f_a] =I_a (x, P_a),
\end{equation}
where $J_a (x, P_a) $ - a collision integral:
$$
I_a(x,P_a)=-\sum \nu_a \int'_a \delta^4(P_F-P_I) \times$$
\begin{equation}\label{21}
W_{IF}(Z_{IF}-Z_{FI})\prod_{I,F}'dP;
\end{equation}
$$W_{FI}=(2\pi)^4|M_{IF}|^22^{-\sum\nu_A+\sum\nu'_b}-$$
-  transition matrix ($ |M _ {IF} | $ - invariant amplitudes of
scattering);
$$Z_{IF}=\prod\limits_{I}f(P^\alpha_A)\prod\limits_F [1\pm f(P^{\alpha'}_B)] ;$$
$$Z_{FI}=\prod\limits_{I}[1\pm f(P^\alpha_A)]\prod\limits_F f(P^{\alpha'}_B) ,$$
''+'' corresponds to bosons, ''- '' - to fermions; details see in
Ref.\cite {Yukin1}, \cite {Yukin2}.

\subsection{The self-consistent system of equations for particles with
scalar interaction}
On the basis of kinetic theory in Ref.\cite {Yukin3} there was
obtained the self-consistent system of equations describing the
statistical self-gravitating system of particles with scalar
in\-te\-rac\-ti\-on. We shall define according to Ref.\cite
{Yukin3} MET of massive scalar field in most general view (with
conformally invariant component and cubic nonlinearity):
$$
T^{ik}_s=\frac{\epsilon}{8\pi}\left[\frac{4}{3}\Phi^{,i}\Phi^{,k}-\frac{1}{3}g^{ik}\Phi^{,j}\Phi_{,j}
+g^{ik}\mu^2_s\Phi^2+\right.
$$
\begin{equation} \label {22}
\left. + \frac {1} {3} \left (R {ik} - \frac {1} {2} R g {ik}
\right) \Phi^2 - \frac {2} {3} \Phi \Phi {, ik} + \right.
\end{equation}
$$ + \left. \frac {2} {3} g {ik} \Phi \Box \Phi
+ \frac {\lambda} {6} g {ik} \Phi^4 \right], $$
where $ \epsilon = + 1 $ in case of scalar interaction with an
attraction of likely scalar charged particles, $ \epsilon =-1 $ -
for repulsion - of likely charged particles. Also by means of
distribution function we'll define a scalar $ \sigma (x) $ \cite
{Yukin3}:

\begin{equation} \label {23}
\sigma (x) = \sum m_Aq_A \int \limits _ {P (x)} dP_A F_A (x, P_A).
\end{equation}
The introduced scalar can be expressed through the spur of MET
particles:
\begin{equation} \label {23}
\sigma (x) = \sum \limits_A \frac {q_A T^A_p} {m_A+q_A \Phi},
\end{equation}
where $T_p $ - spur of MET particles. Further we shall term $
\sigma (x) $ as a {\it scalar density of charges}. It is necessary
to note, that the scalar density of charges is
unam\-bi\-gu\-ous\-ly determined  by conservation laws. Thus,
Einstein's equations for statistical systems of scalar charged
particles take form:
\begin{equation}\label{24}
R^{ik}-\frac{1}{2}R g^{ik}=8\pi( T^{ik}_p+T^{ik}_s),
\end{equation}
and the equation of a scalar field with a source (\ref {23}) is:
\begin{equation} \label {25}
\Box \Phi + \mu^2_s \Phi - \frac {1} {6} R \Phi + \frac {\lambda}
{3} \Phi^3 =-4 \pi \epsilon \sigma.
\end{equation}
The system of equations (\ref {20}), (\ref {24}), (\ref {25})
together with definitions (\ref {15}), (\ref {22}) and (\ref {23})
represent the required closed system of the self-consistent
equations, describing the statistical system of particles with
scalar interaction.

\subsection{The local thermodynamic equilibrium}
The local thermodynamic equilibrium (LTE) is reached in
statistical system, when  a medial run length (time between
collisions) is much less than the typical scale of inhomogeneity
of system (the typical time of evolution). In that cases the
integral of collisions is a main term in the kinetic equations
that leads to so-called functional equations of Boltzmann, having
as the solutions a locally-equilibrium distribution functions:
\begin{equation}\label{26}
f^0(x,P)=\left\{ \exp\left[\frac{-\mu_a+(v,P_a)}{\theta}\right]\mp
1\right\}^{-1},
\end{equation}
where the upper sign corresponds to bosons, lower sign - to
fermions, $ \theta (x) $ - local temperature, identical to all
sorts of particles, $v^i $ - identity timelike macroscopic
velocity vector of statistical system, $ \mu_a (x) $ - chemical
potentials, satisfying the system of linear algebraic equations of
a chemical equilibrium (according to responses (\ref {20-1}):
\begin{equation}\label{27}
\sum\limits_{B=1}^{m'}\nu'_B \mu_B = \sum\limits_{A=1}^m \nu_A
\mu_A.
\end{equation}
Macroscopic characteristics $ \theta (x) $, $v^i (x) $, $ \mu_a
(x) $ are defined from a self-consistent system of equations (\ref
{24}), (\ref {25}), (\ref {27}) and definitions (\ref {15}), (\ref
{22}) and (\ref {23}). Thus we shall obtain \cite {Yu2}:

$$n^i(x)=v^i \frac{2S+1}{2\pi^2}\int\limits_0^\infty
p^2dp\times$$
\begin{equation}\label{28}
\times\left\{\exp\left[\frac{-\mu_a+\sqrt{m_*^2+p^2}}{\theta}\right]\mp
1\right\}^{-1}
\end{equation}
\begin{equation}\label{29}
T_a^{ik}(x)=[{\cal E}_a(x)+P_a(x)]v^iv^k-P_a(x)g^{ik}$$
\end{equation}
where introduced: scalar density of energy
$$ {\cal E}= \sum \limits_a {\cal E} _a $$
and pressure of system of particles:
$$ \quad P = \sum \limits_a P_a: $$

$$ {\cal E} _a (x) = \frac {2S+1} {2 \pi^2} \int \limits_0 ^ \infty
dp\cdot p^2 \sqrt{m^2_*+p^2}\times$$
\begin{equation}\label{30}
\times\left\{\exp\left[\frac{-\mu_a+\sqrt{m_*^2+p^2}}{\theta}\right]\mp1\right\}^{-1},
\end{equation}
$$P_a(x)= \frac{2S+1}{6\pi^2}\int\limits_0^\infty\frac{dp\cdot p^4}{\sqrt{m^2_*+p^2}} \times$$
\begin{equation}\label{31}
\times\left\{\exp\left[\frac{-\mu_a+\sqrt{m_*^2+p^2}}{\theta}\right]\mp1\right\}^{-1}.
\end{equation}
Let's also define an equilibrium density of a scalar charge:
$$ \sigma (x) = \sum \limits \frac {q (2S+1) m_ *} {2 \pi^2} \int \limits_0 ^ \infty
\frac {dp \cdot p^2} {\sqrt {m^2_ * + p^2}} \times $$
\begin{equation}\label{32}
\times\left\{\exp\left[\frac{-\mu_a+\sqrt{m_*^2+p^2}}{\theta}\right]\mp
1\right\}^{-1}.
\end{equation}
The differential consequence of the Einstein equations in the case
of LTE are equations of ideal hydrodynamics, which in the case of
scalar interaction can be lead to the form \cite {Yu2}:
\begin{equation} \label {32a}
\nabla_i [{\cal E} +P) v^i] - (P, _i + \sigma \Phi, _i) v^i=0,
\end{equation}
- the continuity equation for energy, ($ \nabla_i $ -  symbol
covariant derivative) and
\begin{equation}\label{32a}
\nabla_i[{\cal E}+P)v^i]- (P,_i+\sigma \Phi,_i)v^i=0,
\end{equation}
- equations of motion.

Besides, if any vectorial charges conserve (for example,
electrical), we have the additional equations of continuities for
corresponding currents:
\begin{equation} \label {32c}
\nabla_i (v^i \sum \limits_a e_a n_a) =0,
\end{equation}
where $e_a $ - corresponding vector charges.

\section{Completely degenerate Fermi-gas with scalar interaction}
In this article we shall consider completely degenerate
one-sortable Fermi-gas consisting from massive particles with a
spin $1/2 $ as a concrete statistical system. The full
degeneration condition supposes
\begin{equation} \label {33}
\frac {\mu} {\theta} \to \infty.
\end{equation}
In this case the locally-equilibrium distribution function has a
form:
\begin{equation}\label{34}
f^0(x,P)=\left\{\begin{array}{ll} 0,& \mu\leq\sqrt{m_*^2+p^2};\\
 & \\
1,& \mu>\sqrt{m_*^2+p^2}\\
\end{array}\right..
\end{equation}
Therefore the integration of macroscopic densities is
representable in elementary functions:
$${\cal E} =\frac{m_*^4}{8\pi^2}
\left[\psi\sqrt{1+\psi^2}(1+2\psi^2)-\ln
(\psi+\sqrt{1+\psi^2})\right];
$$
$$P =\frac{m_*^4}{24\pi^2}
\left[\psi\sqrt{1+\psi^2}(2\psi^2-3)+3\ln
(\psi+\sqrt{1+\psi^2})\right];
$$
$$
T={\cal E}-3P=\frac{m_*^4}{2\pi^2}\left[\psi\sqrt{1+\psi^2}-\ln
(\psi+\sqrt{1+\psi^2})\right],
$$
\begin{equation}\label{35a}
{\cal E}+P=\frac{m_*^4}{3\pi^2}\psi^3\sqrt{1+\psi^2},
\end{equation}
\begin{equation}\label{35}
\sigma=\frac{q\cdot m_*^3}{2\pi^2}\left[\psi\sqrt{1+\psi^2}-\ln
(\psi+\sqrt{1+\psi^2})\right],
\end{equation}
where $ \psi=p_F/|m_*|$ - the ratio of Fermi momentum to effective
mass. In that case the self-consistent equation of massive scalar
field becomes [1]: $$\Box\Phi+\mu^2\Phi=-\frac{4\pi}{(m+q\Phi)^2}
q T,$$ and a density of number of fermions with Fermi momentum is
connected by relationship Ref.\cite{Land}:
\begin{equation}\label{37}
n(x)=\frac{p^3_F}{3\pi^2}\Rightarrow p_F=(3\pi^2
n(x))^{\frac{1}{3}}.
\end{equation}
Thus, the variable $ \xi $ can be expressed through two scalars -
density of particles number in the natural frame of reference and
scalar potential:
\begin{equation} \label{38}
\psi=\frac{(3\pi^2n(x))^{\frac{1}{3}}}{|m+q\Phi|}.
\end{equation}
\section{Static degenerated Fermi-system}
Let's consider at first a global thermodynamic equilibrium of rest
degenerate gas, which is possible at (Ref.\cite {Yukin3}, \cite
{Eq}):
\begin{equation}\label{39}
\stackunder{\xi}{L}g_{ik}=0;\qquad \stackunder{\xi}{\Phi}=0,
\end{equation}
and $ \xi^i $ is - timelike vector. In that case equilibrium
distribution functions are exactly solutions of kinetic equations,
where it is necessary to specialize a coordinate association of
field magnitudes and ther\-mo\-dy\-na\-mic parameters (Ref.\cite
{Yukin3}):
\begin{equation}\label{40}
v^i=\frac{\xi^i}{\sqrt{\xi,\xi}};  \quad
\theta=\theta_0\sqrt{\xi,\xi}, \quad \mu=\mu_0\sqrt{\xi,\xi}.
\end{equation}
Supposing $$\xi^i = \delta^i_4,$$ we shall obtain:
\begin{equation}\label{41}
\partial_t g_{ik}=0;\quad \partial_t\Phi=0;\quad
\mu=\mu_0\sqrt{g_{44}},
\end{equation}
Let further
$$g_{44}(\infty)=1, \qquad \Phi(\infty)=\Phi_0,\quad
p_F(\infty)=p^0_F.$$
That way we will find:
\begin{equation}\label{42}
\psi=\frac{\sqrt{\rho^2-(1+q\Phi/m)^2}}{(1+q\Phi/m)},
\end{equation}
where the dimensionless function is introduced
$$\rho=\frac{\sqrt{g_{44}(m^2_0+(p_F^0)^2)}}{m_0}\geq 1$$
and
$$m_0=m+q\Phi_0.$$

Introducing further a new dimensionless field variable $\xi$, so
that:
\begin{equation}\label{43}
1+\frac{q\Phi}{m}=\rho\xi\Rightarrow q\Phi=m(\rho\xi-1)\Rightarrow
\psi=\frac{\sqrt{1-\xi^2}}{|\xi|},
\end{equation}

we will obtain:
$${\cal E}_f=$$
\begin{equation}\label{44}\frac{m^4\rho^4}{8\pi^2}\left[(2-\xi^2)\sqrt{1-\xi^2}-\xi^4\ln
\frac{\sqrt{1-\xi^2}+1}{|\xi|} \right];
\end{equation}
$$P_f=$$
\begin{equation}\label{45}\frac{m^4\rho^4}{24\pi^2}\left[(2-5\xi^2)\sqrt{1-\xi^2}-\xi^4\ln
\frac{\sqrt{1-\xi^2}+1}{|\xi|} \right];
\end{equation}
\begin{equation}\label{46}
\sigma=q\frac{m^3\rho^3}{2\pi^2}\left[\sqrt{1-\xi^2}-\xi^2\ln
\frac{\sqrt{1-\xi^2}+1}{|\xi|} \right].
\end{equation}
It is easy to obtain the limiting relations for these
mag\-ni\-tu\-des in case of
\begin{equation}\label{47}
1+\frac{q\Phi}{m}=0 \Rightarrow \xi=0:
\end{equation}
\begin{equation}\label{48_0}
\lim\limits_{\xi\to 0}{\cal E}_f=\frac{m^4\rho^4}{4\pi^2};
\end{equation}
\begin{equation}\label{48}
\lim\limits_{\xi\to
0}P_f=\frac{m^4\rho^4}{12\pi^2}=\frac{1}{3}{\cal E};
\end{equation}
\begin{equation}\label{49}
\lim\limits_{\xi\to 0}\sigma=q\frac{m^3\rho^3}{2\pi^2},
\end{equation}
Thus, in a limit (\ref {47}) system of degenerate fermions becomes
ultrarelativistic.

At the given gravitational field macroscopic
cha\-ra\-c\-te\-ris\-tics of plasma appear explicitly depending
only from the potential scalar field, and the self-consistent
equations of globally equilibrium completely degenerate fermion
system with scalar interaction are described only by one
self-consistent
 equation of type of Klein-Gordon:
$$
\Box\Phi +\mu^2_s\Phi -\frac{1}{6}R\Phi +\frac{\lambda}{3}\Phi^3
+\frac{R}{6}\Phi =$$
\begin{equation}\label{50}
-4\pi\epsilon
q\frac{m^3\rho^3}{2\pi^2}\left[\sqrt{1-\xi^2}-\xi^2\ln
\frac{\sqrt{1-\xi^2}+1}{|\xi|} \right]
\end{equation}
In this case the density of potential energy of scalar fields at
the lack of substance is equal:
\begin{equation}\label{51}
{\cal E}_s=\mu^2\Phi^2+\frac{\lambda}{3}\Phi^4.
\end{equation}
The interaction energy of scalar field with a substance is
obtained from density of the energy of Fermi-system minus the same
magnitude at the lack of a scalar field:
$$
{\cal E}_{in}={\cal E}(\Phi)-{\cal E}(0)=$$
$$=\frac{m^4\rho^4}{8\pi^2}\left[(2-\xi^2)\sqrt{1-\xi^2}-\xi^4\ln
\frac{\sqrt{1-\xi^2}+1}{|\xi|}-\right.$$
\begin{equation}\label{52}\left.-\left(\frac{(2\rho^2-1)\sqrt{\rho^2-1}}{\rho^4}-\frac{1}{\rho^4}\ln
\sqrt{\rho^2-1}+\rho \right) \right]
\end{equation}

Let's discard for simplicity conformally invariant term in the
momentum energy tensor  of scalar field and suppose $ \lambda=0 $;
then the summarized potential energy of scalar field with its
interaction's energy with Fermi-system is equal:
$${\cal E}_\Sigma={\cal E}_S+{\cal E}_{in}=
q^2m^4\rho^2\left\{\alpha^2(\rho\xi-1)^2 +\right.
$$
$$+\frac{1}{8\pi^2}\left[(2-\xi^2)\sqrt{1-\xi^2}-\xi^4\ln
\frac{\sqrt{1-\xi^2}+1}{|\xi|}-\right.$$
\begin{equation}\label{53}\left.\left.-\left(\frac{(2\rho^2-1)\sqrt{\rho^2-1}}{\rho^4}-\frac{1}{\rho^4}\ln
\sqrt{\rho^2-1}+\rho \right) \right]\right\},
\end{equation}
where
$$\alpha^2=\frac{\mu^2}{q^2m^4\rho^2}.$$

In Fig. 1 graphs of  potential energy  ${\cal
E}_\Sigma/q^2m^4\rho^2$ are shown.

\noindent \epsfig {file=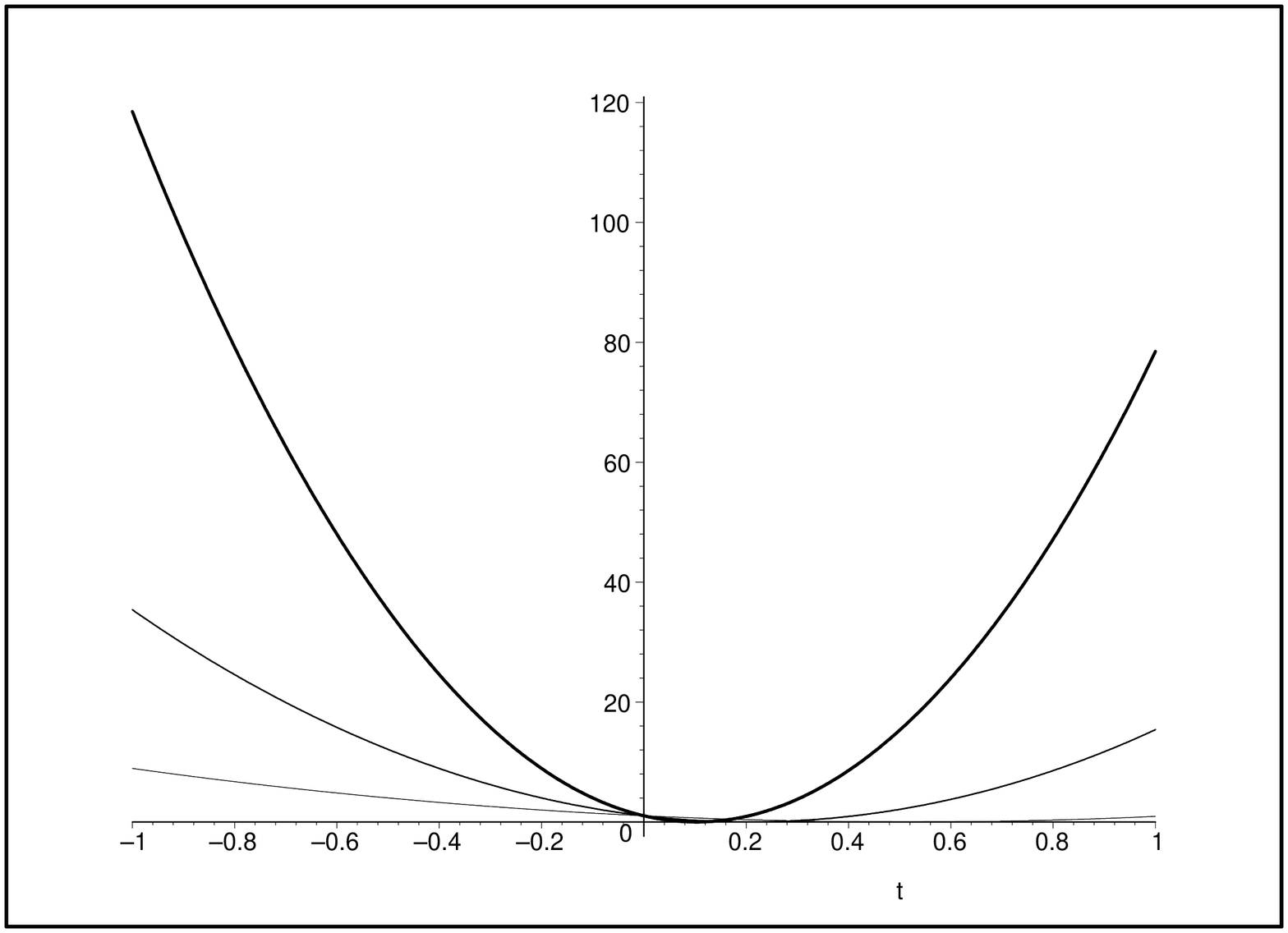,height=8.5cm,width=7cm,angle=-90}
\label {n_evol} \vskip 12pt \noindent {Figure 1:  Dependence graph
of energy $ {\cal E} _ \Sigma/q^2m^4 \rho^2.$ from parameter
$\rho$: (a thin line - $ \rho=2 $, a medial line - $ \rho=5 $,
bold line - $ \rho=10 $); everywhere $ \alpha=1 $. \hfill}%
\vskip 12pt
In Fig. 2 the same graphs in a fine scale are shown - here it is
visible that all graphs intersect a line of zero-point energy, and
their minimums lay in the subzero region.

\noindent \epsfig{file=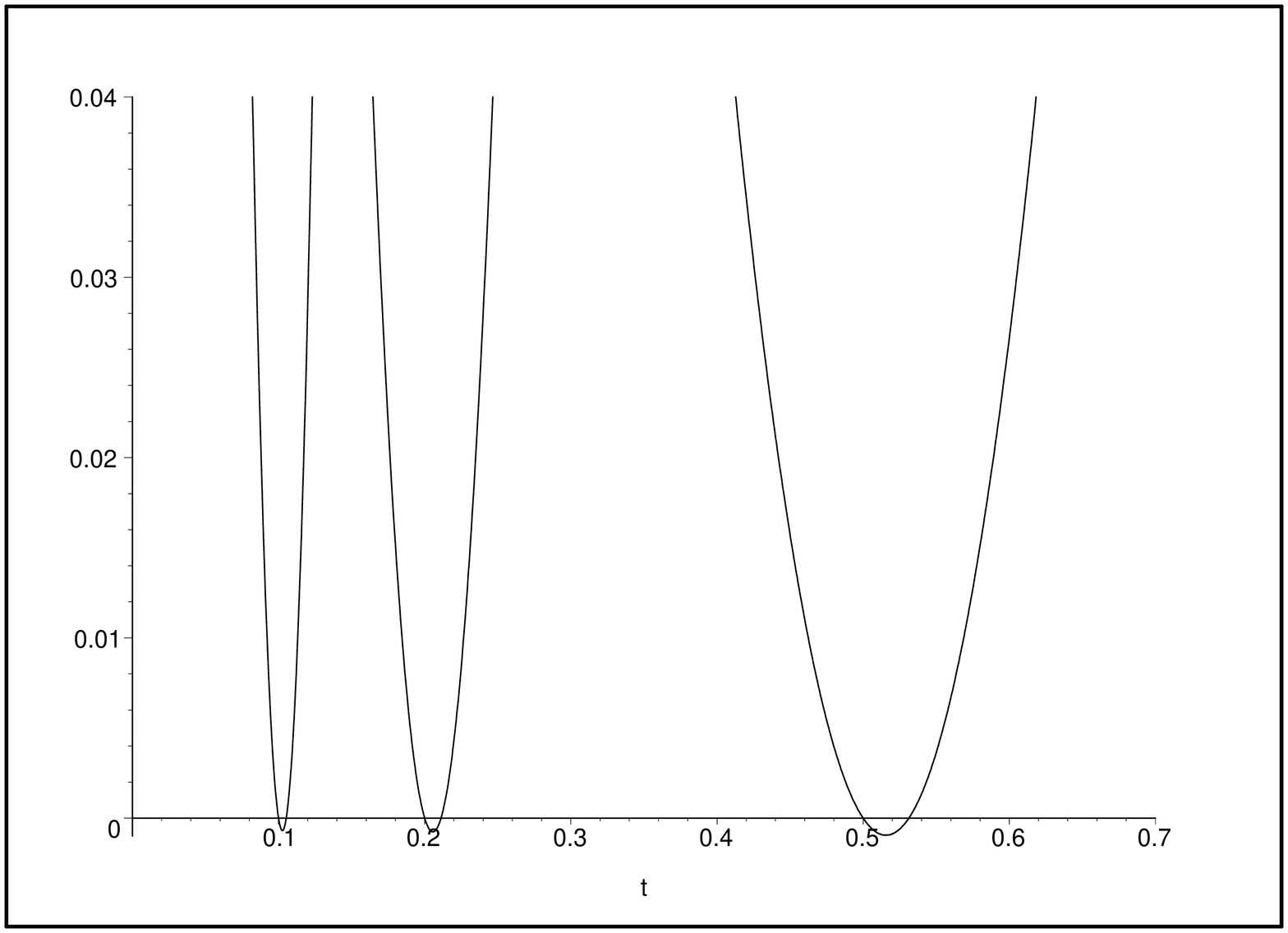,height=8.5cm,width=7cm,angle=-90}
\label {n_evol}\vskip 12pt\noindent {Figure 2: Dependence graph
 of  energy ${\cal E}_\Sigma/q^2m^4\rho^2.$ from parameter
$\rho$=2,5,10(from left to right); everywhere $\alpha=1$. \hfill}
\vskip 12pt
Hence, the account of a scalar field's influence on the
statistical system leads to the appearance of non-linear effective
potential of interaction which can have minimums at nonzero values
$ \Phi $, that can perform excessive an operation of artificial
introduction of cubic nonlinearity in dynamic equations of scalar
field. Besides, the introduction of scalar interactions leads to
the effective mechanism of state equations regulation, which can
appear important in cosmological situation.

\section{Homogeneous degenerate Fermi-system in cosmology}
Let's consider a cosmological situation when the substance is
presented only as degenerate Fermi-system with scalar interaction
of particles. In this case the self-consistent system of Einstein
and Klein-Gordon equations  with a scalar source in the metric
\begin{equation}\label{54}
ds^2=dt^2-a^2(t)(dx^2+dy^2+dz^2)
\end{equation}
takes form:
\begin{equation}\label{55}
3\frac{\dot{a}^2}{a^2}=8\pi
\end{equation}
In this metric $$v^i=\delta^i_4$$ and from the relation (\ref
{37}) and conservation law of particles:
$$\partial_i \sqrt{-g}n v^i=0$$
we will obtain the momentum integral instead of the energy
integral (\ref {41}):
\begin{equation}\label{56}
a p_F=\mbox{Const}.
\end{equation}
Supposing that
$$\Phi=\Phi(t); \Rightarrow {\cal E}={\cal E}(t);\; P=P(t),$$
we shall obtain the structure of summary momentum-energy tensor of
scalar field in the form of  momentum-energy tensor of ideal fluid
with macroscopic velocity $v^i $ and a density of energy $ {\cal
E} _S $ and pressure $P_S $:
\begin{equation}\label{57}
{\cal E}_s=\frac{1}{8\pi}(\dot{\Phi}^2+\mu^2_s\Phi^2);\quad
P_s=\frac{1}{8\pi}(\frac{1}{3}\dot{\Phi}^2-\mu^2_s\Phi^2).
\end{equation}
As it's known, (see, for example, \cite {Land1}), in the metric
(\ref {54}) Einstein independent equations have a form:
\begin{equation}\label{58}
3\frac{\dot{a}^2}{a^2}=8\pi\varepsilon;
\end{equation}
\begin{equation}\label{59}
3\frac{\dot{a}}{a}=-\frac{\dot{\varepsilon}}{\varepsilon+P}.
\end{equation}
According to the momentum integral (\ref {56}) let's introduce new
dimensionless variables and parameters:
\begin{equation}\label{60}
\varphi=\frac{1}{\psi}; \quad \beta=\frac{p^0_F}{m}.
\end{equation}
Then:
\begin{equation}\label{61}
\Phi=\frac{m}{q}\bigl(\frac{\beta\varphi}{a}-1\bigr), \quad
m_*=m\beta\frac{\varphi}{a},
\end{equation}
where we have supposed:
$$ p^0_F=p_F (t_0); a (t_0) =1. $$
Thus, it remains two unknown functions, $a (t) $ and $ \psi (t) $.

As the equation of a scalar field is a consequence of  Einstein
equations  it can be missed.
Considering an explicit structure of expressions for a density of
energy and pressure,let's introduce for convenience conformal
densities according to the rule:
\begin{equation}
X^*=a^4X.
\end{equation}
Einstein equations for conformal densities have a form:
\begin{equation}\label{62}
{\cal E}^*\ '-\frac{a'}{a}({\cal E}^*-3P^*)=0;
\end{equation}
\begin{equation}\label{63}
3a'^2=8\pi {\cal E}^*,
\end{equation}
where we have changed over to the temporal variable derivation  $
\eta $ by rule:
$$ad\eta=dt \Rightarrow \dot{X}=\frac{X'}{a};$$
- $X'=dX/d \eta $. In new variables densities of energy and
pressure of Fermi-system and scalar field can be written in the
form of:
$$
{\cal E}_f^*=\frac{m^4\beta^4}{8\pi^2}\times
$$
\begin{equation}\label{64}
\times\left[
\sqrt{1+\varphi^2}(2+\varphi^2)-\varphi^4\ln\frac{1+\sqrt{1+\varphi^2}}{\varphi}\right];
\end{equation}
$${\cal E}_f^*-3P_f^*=$$
\begin{equation}\label{65}\frac{m^4\beta^4}{3\pi^2}\varphi^2\bigl(\sqrt{1+\varphi^2}-
\varphi^2\ln\frac{1+\sqrt{1+\varphi^2}}{\varphi}\bigr);
\end{equation}
\begin{equation}\label{66}
{\cal E}^*_s=\frac{\beta^2m^2}{8\pi
q^2}\left[\varphi^2-2\varphi\varphi'+\varphi'a+\mu^2a^2\Bigl(\varphi-\frac{a}{\beta}\Bigr)\right];
\end{equation}
\begin{equation}\label{67}
{\cal
E}^*_s-3P^*_s=\frac{\mu^2_s\Phi^2}{2\pi}=\frac{\beta^2\mu^2_s
m^2}{2\pi q^2}a^2\Bigl(\varphi-\frac{a}{\beta}\Bigr)^2.
\end{equation}
Calculating, we'll find:
$$\varphi\frac{d {\cal E}^*_f}{d\varphi}=\frac{m^4\beta^4}{2\pi^2}\varphi^2\times$$
$$\left[\sqrt{1+\varphi^2}-\varphi^2\ln\Bigl(\frac{1+\sqrt{1+\varphi^2}}{\varphi} \Bigr) \right].$$

\begin{equation} \label{68}
\varphi\frac{d{\cal E}^*_f}{d\varphi}=\frac{3}{2}({\cal
E}_f^*-3P^*_f)
\end{equation}
Substituting the obtained expressions in the equations (\ref{66}),
(\ref{67}), we will obtain the system of two ordinary differential
equations of the first and second order concerning variables
$a(\eta)$ and $\varphi(\eta) $. Below some results of numerical
integration of this system are represented.

\noindent \epsfig{file=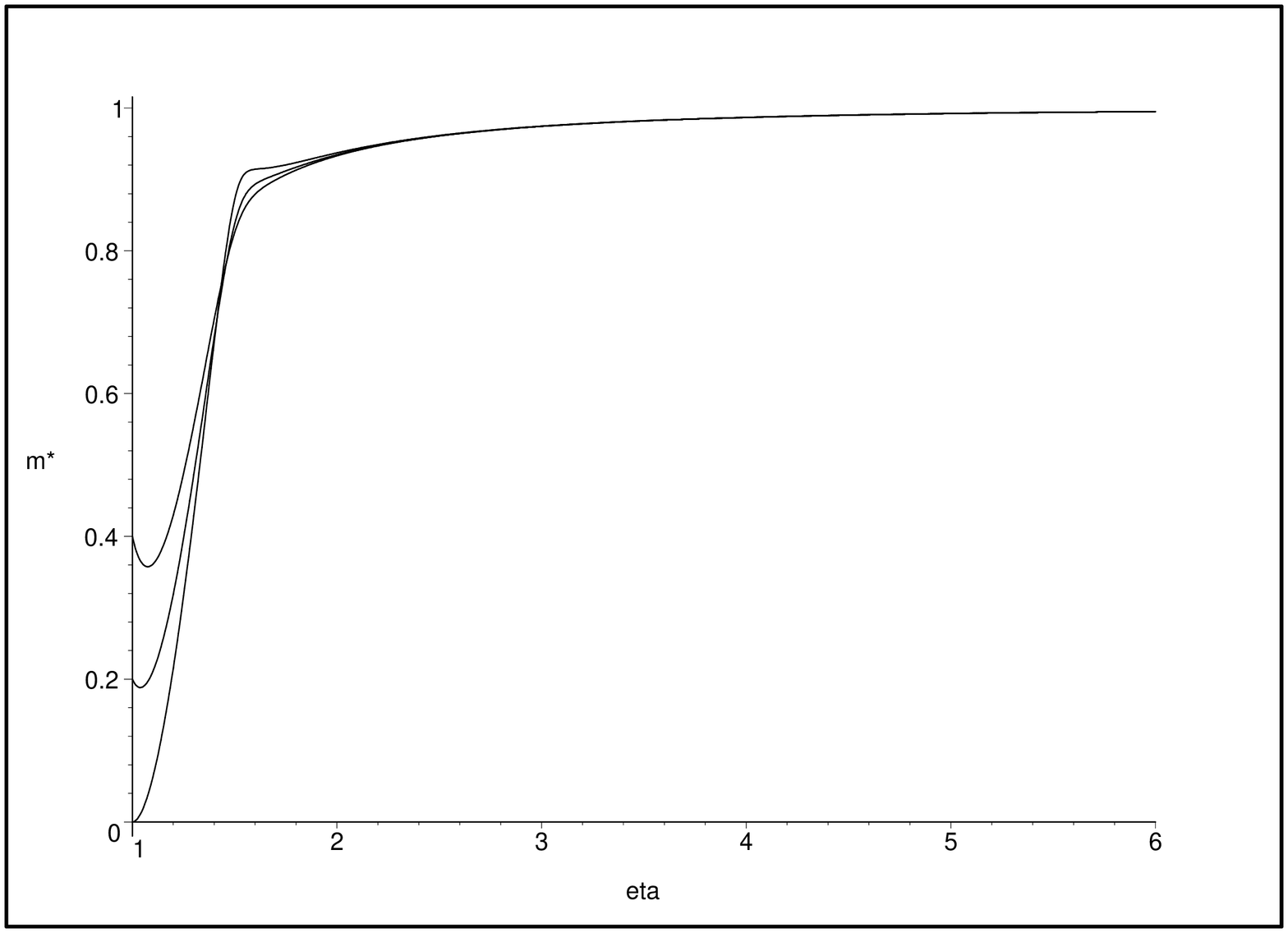,
height=8.5cm,width=6cm,angle=-90}\label{gm1} \vskip 12pt \noindent
Figure\ 3:\hskip 12pt{\sl Evolution of the effective mass $m_*$ of
fermions depending on initial mass.\hfill} \vskip 0pt
\noindent \epsfig {file=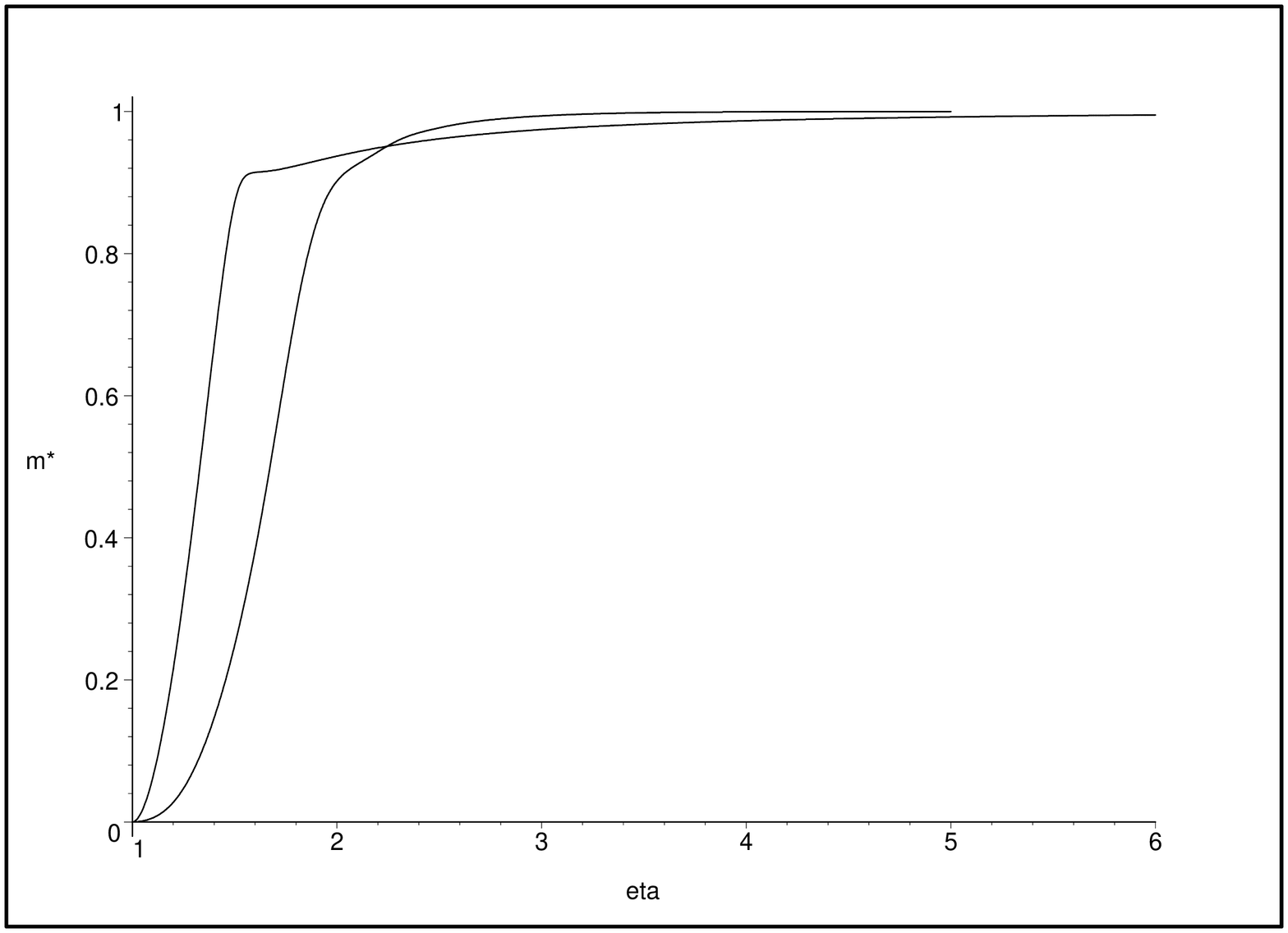, height=8.5cm, width=6cm,
angle =-90} \label {gm_beta} \vskip 20pt \noindent Figure\ 4:
\hskip 12pt {\sl Evolution of the effective mass $m_ * $ fermions
depending on parameter $ \rho $ = 2-thin line and $ \beta $ = 10 -
bold line. \hfill} \vskip 0pt
\noindent\epsfig{file=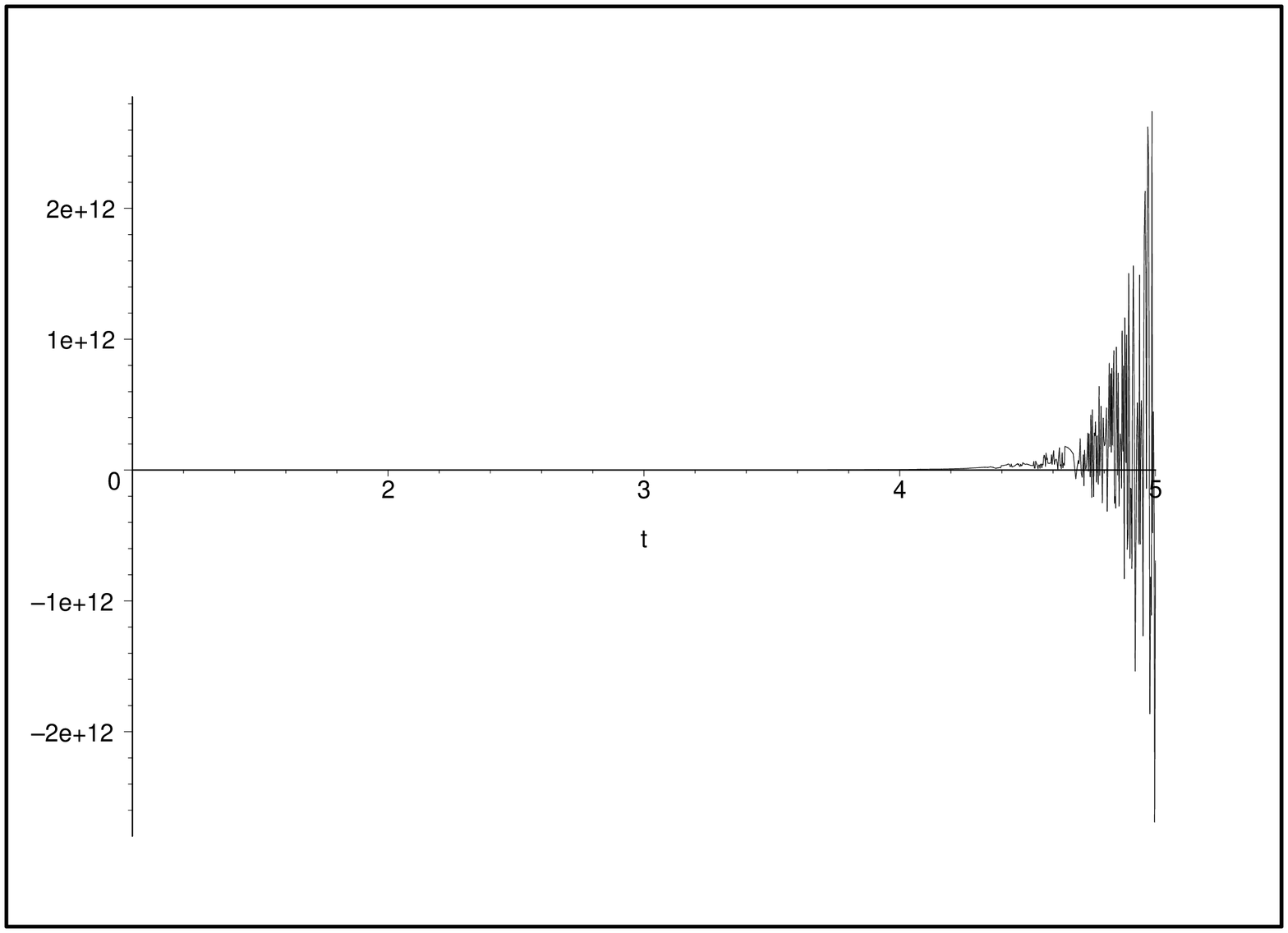,height=8.5cm,width=6cm,angle =-90}
\label{sing}\vskip 12pt \noindent Figure\ 5: \hskip 12pt {\sl
Evolution of a track of conformal MET ${\cal E}^*-3P^*$. The
singularity is visible near $\eta \sim 5$.\hfill}\vskip 0pt

\section*{Conclusion}
Mentioned above results of numerical integration of Einstein -
Klein - Gordon equations for degenerate Fermi-gas of scalar
charged particles exhibit an essential influence of Fermi-system
interaction with scalar field on the summary equation system
state. Detection of this influence by means of numerical modelling
of system will be considered in our following article.

\end{document}